\documentclass[11pt,a4paper]{amsart}

\usepackage{amssymb,amsfonts,amsmath}
\usepackage[latin1]{inputenc}

\newtheorem{theorem}{Theorem}
\newtheorem{proposition}[theorem]{Proposition}
\newtheorem{lemma}[theorem]{Lemma}
\newtheorem{corollary}[theorem]{Corollary}

\newtheorem{remark}[theorem]{Remark}
\newtheorem{definition}[theorem]{Definition}

\def\handle{\mathop{\rm handle}\nolimits}
\def\Core{\mathop{\rm Core}\nolimits}
\def\Even{\mathop{\rm Even}\nolimits}

\makeatletter
\@namedef{subjclassname@2010}{%
\textup{2010} Mathematics Subject Classification}
\makeatother

\begin{document}

\title[]{Extended core and choosability of a graph}
\author{Yves Aubry, Jean-Christophe Godin and Olivier Togni}
\address{Institut de Math\'ematiques de Toulon, Universit\'e du Sud Toulon-Var,  France\\
and Laboratoire LE2I, Universit\'e de Bourgogne, France}

\email{yves.aubry@univ-tln.fr,  godinjeanchri@yahoo.fr\\
and olivier.togni@u-bourgogne.fr}



\date{\today}

\begin{abstract}
A graph $G$ is $(a,b)$-choosable if for any color list of size $a$ associated with each vertices, one can choose a subset of $b$ 
colors such that adjacent vertices are colored with disjoint color sets.
This paper shows an equivalence between the $(a,b)$-choosability of a graph and the $(a,b)$-choosability of one of its subgraphs 
called the extended core. As an application, this result allows to prove the $(5,2)$-choosability and $(7,3)$-colorability of 
triangle-free induced subgraphs of the triangular lattice.
\end{abstract}

\maketitle



\section{Introduction}

Let $G=(V(G),E(G))$ be a graph  where $V(G)$ is the set of vertices and $E(G)$ is the set of edges, and let $a$, $b$, $n$ and $e$ be integers. 

Given a list $L$ of $G$ {\em i.e.} a map $L : V(G) \rightarrow \mathcal{P}({\mathbb N})$ and a weight $\omega$
of $G$ {\em i.e.} a map $\omega : V(G) \rightarrow {\mathbb N}$, an {\em $(L,\omega)$-choosability} $c$ of $G$ is a list 
of the weighted graph $G$ such that for all  $vv' \in E(G)$: 
$$c(v) \subset L(v),\ \ \  | c(v) | =\omega(v) \ \ \ {\rm and}\ \ \   c(v) \cap c(v')  = \emptyset.$$
We say that  $G$ is {\em $(L,\omega)$-choosable} if there exists an $(L,\omega)$-choosability $c$ of $G$.
An $(L,b)$-choosability $c$ of $G$ is an $(L,\omega)$-choosability of $G$ such that 
for all $v \in V(G)$, we have $\omega(v)=b .$
A $a$-list $L$ of $G$ is a list of $G$ such that for all $v \in V(G)$, we have  $|L(v) |=a$.
The graph $G$ is said to be {\em $(a,b)$-choosable} if for any $a$-list $L$ of $G$, there exists an $(L,b)$-choosability $c$ of $G$.
If the graph is $(a,b)$-choosable for the $a$-list $L$ such that $L(v)=L(v')$ for all vertices $v,v'$, then $G$ is $(a,b)$-colorable.

The concept of list coloring and choosability was introduced by Vizing~\cite{viz76} and independently by Erd\H{o}s, Rubin and Taylor~\cite{erdo79}.
Since then, it has been the subject of many works (see~\cite{tom94,gravier96,voi96,borodinkostochkawoodall97,voi98} and \cite{GutnerTarsi,hav09} for more recent papers).
In order to characterize $2$-choosable ({\em i.e.} $(2,1)$-choosable) graphs, Erd\H{o}s et al. defined the notion of the core of a graph.
The aim of this paper is to extend the notion of core to obtain some characterizations of $(a,b)$-choosable graphs. 
Basically, extended cores will be obtained by removing vertices of low degree and induced paths with conditions on the vertex degrees, called handles.
Using results on the choosability of a weighted path~\cite{nous} and extending some of them (Section~\ref{path}), 
the main result is Theorem~\ref{theoremcorech} of Section 3 that shows the equivalence between the $(a,b)$-choosability of a graph $G$ and 
the $(a,b)$-choosability of a subgraph of $G$ called its first extended core.
Some applications of this theorem for triangle-free induced subgraphs of the triangular lattice are given in Section 4 and 5, 
where it is shown that these graphs are $(5,2)$-choosable (Theorem~\ref{theoreme55these}) and $(7,3)$-colorable (Theorem~\ref{havet73}), 
thus giving another proof of Havet's result~\cite{hav01}.

\medskip
In order to extend a result on the choosability of a weighted path, let us first give some definitions and some known results.

The \textsl{path} $P_{n+1}$ of length $n$ is the graph with vertex set $V=\{v_{0},v_{1},\dots,v_{n}\}$ and edge set 
$E=\bigcup_{i=0}^{n-1} \{v_{i}v_{i+1}\}$. To simplify the notations $L(i)$ denotes $L(v_{i})$ and $c(i)$ denotes
$c(v_{i})$.\\ 

\begin{definition} For the path $P_{n+1}$  of length $n$,
\begin{itemize}
 \item a {\em waterfall} list $L$ is a list such that for all $i,j \in \{0,\dots,n\}$ 
with $|i-j| \geq 2$, $L(i) \cap L(j) = \emptyset$;
\item two lists $L$ and $L'$ are {\em similar} if and only if $P_{n+1}$  is $(L,\omega )$-choosable whenever $P_{n+1}$  is $(L',\omega )$-choosable;
\item the {\em amplitude} $A_{i,j}(L)$ (or $A_{i,j}$) of a list $L$ is $A_{i,j}(L)=\cup_{k=i}^{j}L(k)$;
\item a list $L$ is {\em good} if $|L(i)|\geq \omega(i) + \omega (i+1)$ for any $i, 1\leq i\leq n-1$.
\end{itemize}
\end{definition}


We have proved in \cite{nous} the following results:

\begin{proposition}
\label{llf}
 For any good list $L$ of $P_{n+1}$ , there exists a similar waterfall list $L^c$ with $\vert L^c(i)\vert=\vert L(i)\vert$ for all $i\in\{0,\ldots,n\}$.
\end{proposition}

\begin{theorem}
\label{theolistecascadechoisissable}
Let $L^{c}$ be a waterfall list of a weighted path 
 $P_{n+1}$. Then $P_{n+1}$ is $(L^{c},\omega)$-choosable if and only if:
$$\forall i,j \in \{0,\dots,n\} : \: |A(i,j)(L^{c})| \geq \sum_{k=i}^{j} \omega (k) .$$ 
\end{theorem}

\begin{corollary}
\label{lemmeencascadeequige}
Let $L^{c}$ be a good waterfall list of a weighted path $P_{n+1}$ such that $|L^c(n)| \geq \omega(n)$. Then $P_{n+1}$ is  $(L^{c},\omega)$-choosable if and only if  
$$ \forall j \in \{0,\dots,n\}: \:|A(0,j)(L^{c})| \geq \sum_{k=0}^{j} \omega (k). $$ 
\end{corollary}

\begin{corollary}
\label{theorem48these}
Let $L$ be a list of $P_{n+1}$ such that $|L(0)|=|L(n)|= b$,
and $|L(i)|=a=2b+e$ for all $i \in \{1,\dots,n-1\} $. 
$$If \ n \geq \Even\Bigl(\frac{2b}{e}\Bigr) \ then \ P_{n+1} \ is \ (L,b)-choosable$$
where $\Even(x)$ is the smallest even integer $p$ such that $p\geq x$.
\end{corollary}



\section{Choosability of a path}\label{path}

The purpose of this section is to extend the result of Corollary \ref{theorem48these} to other constrained lists, namely to $1$-reduced lists.

\begin{definition}
A list $L$ is said to be a   $1$-reduced list of $P_{n+1}$ if $|L(0)| = b$, for any $i \in \{1,\dots,n-2\} : |L(i)|=a$, $|L(n-1)| =|L(n)|= b + e$, and $|A(n-1,n)(L)| \geq 2b$. 
\end{definition}

\begin{theorem}
\label{theorem49these}
Let $n,a,b,e$ be four integers such that  $a= 2b+e$. Let $L$ be a $1$-reduced list of a path  $P_{n+1}$. 
$$If \ n = \Even(\frac{2b}{e}) \ then \ P_{n+1} \ is \ (L,b)-choosable.$$
\end{theorem}

\begin{proof}
Since
$$|A(n-1,n)|=|L(n-1)|+|L(n) \backslash L(n-1)| = (b+e)+|L(n) \backslash L(n-1)| \geq 2b,$$
we obtain that $|L(n) \backslash L(n-1)| \geq b-e$. Let $D$ be a set such that 
$$D \subset L(n) \backslash L(n-1) \ and \ |D|=b-e.$$
Let $L'$  be the new list constructed with  $L$ such that (see the following figure):
$$L'(i) = \left \{
\begin{array}{l}
L(i) \ \ \  \ \  \ \ if \ i \in \{0,\dots,n-1\} \\
L(n) \backslash D \ \ \ \  \ \ \  \ \ otherwise
\end{array}
\right.
$$

\unitlength=0.8cm
\begin{picture}(20,6.2)
\put(5,2.5){\framebox(0.5,2){}}
\put(5.8,1){\framebox(0.5,2){}}
\put(13,2.5){\framebox(0.5,2){}}
\put(13.8,2.3){\framebox(0.5,0.7){}}
\put(0.1,5.5){$L( )$}
\put(1.8,5.5){0}
\put(2.6,5.5){.}
\put(3.4,5.5){.}
\put(4.2,5.5){.}
\put(4.7,5.5){$n-1$}
\put(6,5.5){$n$}
\put(8.5,5.5){$L'( )$}
\put(9.8,5.5){0}
\put(10.6,5.5){.}
\put(11.4,5.5){.}
\put(12.2,5.5){.}
\put(12.7,5.5){$n-1$}
\put(14,5.5){$n$}
\multiput(1.8,5)(.1,0){24}{\line(1,0){0.05}}
\multiput(1.8,1)(.1,0){24}{\line(1,0){0.05}}
\multiput(1.8,1)(0,0.1){40}{\line(0,1){0.05}}
\multiput(4.2,1)(0,0.1){40}{\line(0,1){0.05}}
\put(1.9,4){Remainder}
\put(2.7,3){of}
\put(2.5,2){lists}
\multiput(9.8,5)(.1,0){24}{\line(1,0){0.05}}
\multiput(9.8,1)(.1,0){24}{\line(1,0){0.05}}
\multiput(9.8,1)(0,0.1){40}{\line(0,1){0.05}}
\multiput(12.2,1)(0,0.1){40}{\line(0,1){0.05}}
\put(9.9,4){Remainder}
\put(10.7,3){of}
\put(10.5,2){lists}
\put(7.5,4.5){\vector(1,0){1}}
\put(8.5,4.5){\vector(-1,0){1}}
\multiput(5.8,1)(.1,0){5}{\line(0,1){1.3}}
\put(5.8,2.3){\line(1,0){.5}}
\put(6.8,2.3){\vector(0,-1){1.3}}
\put(6.8,1){\vector(0,1){1.3}}
\put(7,1.5){$b-e$}
\put(6.8,2.3){\vector(0,1){0.7}}
\put(6.8,3){\vector(0,-1){0.7}}
\put(7,2.5){$2e$}
\put(14.8,2.3){\vector(0,1){0.7}}
\put(14.8,3){\vector(0,-1){0.7}}
\put(15,2.5){$2e$}
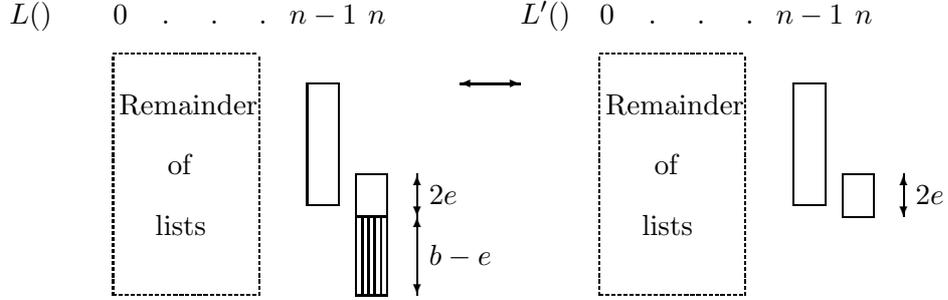
\put(.8,.1){Fig. 1. Construction of $L'$ (on the right) for a $1$-reduced list $L$ (on the left).}
\end{picture}
\\

A new weight function   $\omega'$ is constructed such that:  
$$\omega'(i) = \left \{
\begin{array}{l}
b \ \ \  \ \  \ \ if \ i \in \{0,\dots,n-1\} \\
e \ \ \ \  \ \ \  \ \ otherwise.
\end{array}
\right.
$$
We are going to prove that: 
$$ if \ P_{n+1} \ is \ (L',\omega')-choosable \ then \ P_{n+1} \ is \ (L,b)-choosable.$$
Indeed, if $c'$ is an $(L',\omega')$-choosability of $ P_{n+1}$, then we construct $c$ such that:
$$c(i) = \left \{
\begin{array}{l}
c'(i) \ \ \  \ \  \ \ \ \ if \ i \in \{0,\dots,n-1\} \\
c'(n) \cup D \ \ \ \  \ \ \  \ \ otherwise.
\end{array}
\right.
$$
Since $D \cap L(n-1) = \emptyset$, we have $c(n-1) \cap c(n) = \emptyset$ and then $c$ is an $(L,b)$-choosability of $ P_{n+1}$.\\
Now, this new list   $L'$ is a good list  of $P_{n+1}$ and $|L'(n)| \geq \omega'(n)$. 
Proposition \ref{llf} shows that there exists a waterfall list $L^c$ similar to $L'$ such that for all $k$ we have 
 $|L^c(k)|=|L'(k)|$. 
 
Thanks to Corollary \ref{lemmeencascadeequige}, it remains to check that:
$$ \forall j \in \{0,\dots,n\} : \:|A(0,j)(L^{c})| \geq \sum_{k=0}^{j} \omega' (k). $$

Case 1: $j \in \{0,\dots,n-2\}$. Since the list is a waterfall list, we have: 
$$ |A(0,j)(L^{c})| \geq \left \{
\begin{array}{l}
\sum_{\substack {k=0 \\ k\ even}}^j |L^c(k)|= b+a\frac{j}{2}  \ \  \ \ \ \  \ \ \ \ if  \ j \ is \ even \\
 \sum_{\substack {k=0 \\ k\ odd}}^j |L^c(k)| = a\frac{j+1}{2}   \  \ \ \  \ \ \ \  \ otherwise
\end{array}
\right.
$$
and the weight function satisfies $\sum_{k=0}^{j} \omega' (k)=(j+1)b$. Hence, we deduce that 
$$|A(0,j)(L^{c})| \geq \sum_{k=0}^{j} \omega' (k).$$

Case 2: $j=n-1$. Since the list is a waterfall list, we have: 
$$ |A(0,n-1)(L^{c})| \geq \sum_{\substack {k=0 \\ k\ odd}}^{n-1} |L^c(k)|=(b+e)+a\frac{n-2}{2},$$ 
and $\sum_{k=0}^{n-1} \omega' (k)=nb$. Then $b+e+a\frac{n-2}{2} \geq nb$ if and only if $\frac{en}{2} \geq b$, which is true by hypothesis since  $n = \Even(\frac{2b}{e})$, thus
$$|A(0,n-1)(L^{c})| \geq \sum_{k=0}^{n-1} \omega' (k).$$

Case 3: $j=n$. Since the list is a waterfall list, we have: 
$$ |A(0,n)(L^{c})| \geq  \sum_{\substack {k=0 \\ k\ even}}^n |L^c(k)|=b+2e+a\frac{n-2}{2},$$
and $\sum_{k=0}^{n} \omega' (k)=nb+e$, then $b+2e+a\frac{n-2}{2} \geq nb+e$ if and only if $\frac{en}{2} \geq b$, which is true by hypothesis since $n = \Even(\frac{2b}{e})$, thus
$$|A(0,n)(L^{c})| \geq \sum_{k=0}^{n} \omega' (k).$$
\end{proof}






\section{The extended core of a graph}

The purpose of this section is to prove Theorem \ref{theoremcorech} which gives the equivalence between the $(a,b)$-choosability
 of a graph $G$ and the $(a,b)$-choosability of one of its subgraphs, its first extended core, denoted $\Core_{ch}(x,1)(G)$ (with the idea to take $x=\frac{a}{b}$).

The first extended core   is a generalization of the core, introduced by Erd\H{o}s, Rubin and Taylor in \cite{erdo79} for 2-choosable graphs.

\begin{definition}
A $\handle_x$ of length $n$ in a graph $G$ is a path  $\{v_0,\dots,v_n\}$ such that the vertices  $\{v_1,\dots,v_{n-1}\}$ have degree
less or equal to $\lfloor x \rfloor$ and for all  $i,j \in \{1,\dots,n-1\}$ we have: if $|i-j| \geq 2$ then $v_i v_j
\notin E$. The interior of the $\handle_x$ is the set of vertices
 $\{v_1,\dots,v_{n-1}\}$.
\end{definition}

\unitlength=0.8cm
\begin{picture}(20,4)
\put(3,2){\line(1,0){10}}
\put(3,2){\circle*{0.2}}
\put(4,2){\circle*{0.2}}
\put(5,2){\circle*{0.2}}
\put(11,2){\circle*{0.2}}
\put(12,2){\circle*{0.2}}
\put(13,2){\circle*{0.2}}
\put(2.8,1.2){$v_0$}
\put(3.8,1.2){$v_1$}
\put(4.8,1.2){$v_2$}
\put(10.8,1.2){$v_{n-2}$}
\put(11.8,1.2){$v_{n-1}$}
\put(12.8,1.2){$v_n$}
\put(3.6,2.8){$\lfloor x \rfloor$}
\put(4.6,2.8){$\lfloor x \rfloor$}
\put(10.6,2.8){$\lfloor x \rfloor$}
\put(11.6,2.8){$\lfloor x \rfloor$}
\put(2.7,1.7){\line(1,1){.6}}
\put(3.7,1.7){\line(1,1){.6}}
\put(4.7,1.7){\line(1,1){.6}}
\put(10.7,1.7){\line(1,1){.6}}
\put(11.7,1.7){\line(1,1){.6}}
\put(12.7,1.7){\line(1,1){.6}}
\put(3.3,1.7){\line(-1,1){.6}}
\put(4.3,1.7){\line(-1,1){.6}}
\put(5.3,1.7){\line(-1,1){.6}}
\put(11.3,1.7){\line(-1,1){.6}}
\put(12.3,1.7){\line(-1,1){.6}}
\put(13.3,1.7){\line(-1,1){.6}}
\multiput(6,1.4)(.3,0){15}{\line(1,0){0.05}}
\put(5,.1){Fig. 2. Example of a $\handle_x$.}
\end{picture}
\\

\begin{definition}
A $1$-$\handle_x$ of length $n$ is a $\handle_x$ of length $n$ such that $v_n$ has degree less or equal to  $\lfloor x +1 \rfloor$ and has a neighbor 
 $v_{n+1}$ of degree less or equal to   $\lfloor x \rfloor$ and for all  $i,j \in \{1,\dots,n+1\}$ we have:  if $|i-j|
\geq 2$ then $v_i v_j \notin E$. 
\end{definition}

\unitlength=0.8cm
\begin{picture}(20,4)
\put(3,2){\line(1,0){10}}
\put(3,2){\circle*{0.2}}
\put(4,2){\circle*{0.2}}
\put(5,2){\circle*{0.2}}
\put(11,2){\circle*{0.2}}
\put(12,2){\circle*{0.2}}
\put(13,2){\circle*{0.2}}
\put(14,3){\circle*{0.2}}
\put(2.8,1.2){$v_0$}
\put(3.8,1.2){$v_1$}
\put(4.8,1.2){$v_2$}
\put(10.8,1.2){$v_{n-2}$}
\put(11.8,1.2){$v_{n-1}$}
\put(12.8,1.2){$v_n$}
\put(13.8,2.2){$v_{n+1}$}
\put(3.6,2.8){$\lfloor x \rfloor$}
\put(4.6,2.8){$\lfloor x \rfloor$}
\put(10.6,2.8){$\lfloor x \rfloor$}
\put(11.6,2.8){$\lfloor x \rfloor$}
\put(13.6,.8){$\lfloor x+1 \rfloor $}
\put(13.6,3.8){$\lfloor x \rfloor$}
\put(2.7,1.7){\line(1,1){.6}}
\put(3.7,1.7){\line(1,1){.6}}
\put(4.7,1.7){\line(1,1){.6}}
\put(10.7,1.7){\line(1,1){.6}}
\put(11.7,1.7){\line(1,1){.6}}
\put(12.7,1.7){\line(1,1){.6}}
\put(3.3,1.7){\line(-1,1){.6}}
\put(4.3,1.7){\line(-1,1){.6}}
\put(5.3,1.7){\line(-1,1){.6}}
\put(11.3,1.7){\line(-1,1){.6}}
\put(12.3,1.7){\line(-1,1){.6}}
\put(13.3,1.7){\line(-1,1){.6}}
\multiput(6,1.4)(.3,0){15}{\line(1,0){0.05}}
\put(13,2){\line(1,1){1.3}}
\put(14.3,2.7){\line(-1,1){.6}}
\put(5,.1){Fig. 3. Example of a $1$-$\handle_x$.}
\end{picture}
\\

We define the first extended core, denoted  $\Core_{ch}(x,1)(G)$, of a graph $G$ as the induced subgraph of $G$ obtained inductively when we remove:
\begin{itemize}
\item its vertices of degree  $0,1,\dots,\lfloor x-1 \rfloor$, 
\item the interior of its  $\handle_x$ of length $n \geq \Even(\frac{2}{x-\lfloor x \rfloor})$,
\item  the interior of its $1$-$\handle_x$ of length $n = \Even(\frac{2}{x-\lfloor x \rfloor}) -1$.
\end{itemize}

Let us remark that the definition of $\Core_{ch}(x,1)(G)$ does not depend on the order which we use to remove the
vertices and that the subscript 'ch' means that we will use it for choosability purpose.\\

\begin{theorem}
\label{theoremcorech}
Let $a,b$ be two integers and $x$ be a rational number such that   $\frac{a}{b} \geq x$. For any graph $G$, we have the following equivalence:

$$G \ is \ (a,b)-choosable \Leftrightarrow \Core_{ch}(x,1)(G) \ is \ (a,b)-choosable.$$
\end{theorem}

\begin{proof}
Since $\Core_{ch}(x,1)(G)$ is a subgraph of $G$, if $G$ is $(a,b)$-choosable then $\Core_{ch}(x,1)(G)$ is of course $(a,b)$-choosable.

Conversely, suppose that $\Core_{ch}(x,1)(G)$ is $(a,b)$-choosable. Let us discuss the different cases and let $e$ be the integer defined by:

$$e=a-\lfloor x \rfloor b \geq 0 \ .$$ 

Case 1: we remove a vertex of low degree. Let $v \in V(G)$ such that its degree $d(v) \leq \lfloor x \rfloor -1$. 
Let $N(v)$ be the set of neighbors of $v$ in $G$. Suppose that $G-\{v\}$ is $(a,b)$-choosable and let $L$ be an $a$-list of $G$. 
Then there exists an $(L,b)$-choosability $c$ of  $G-\{v\}$. 

Let $L'(v)=L(v) \setminus \bigcup_{w \in N(v)} c(w)$. Then 
$$|L'(v)| \geq |L(v)| - d(v)b \geq a - (\lfloor x \rfloor - 1)b=b+e \geq b.$$
Hence, we can complete the choosability with 
 $c(v) \subset L'(v)$ such that $|c(v)|=b$, and thus $G$ is $(a,b)$-choosable.\\

Case 2: $x$ is an integer  ($x= \lfloor x \rfloor$). Since $\Even(\frac{2}{x-\lfloor x \rfloor})= \infty$, the extended core $\Core_{ch}(x,1)(G)$ is limited to the case 1, and  the proof is done.\\

Case 3: $x$ is not an integer  ($x > \lfloor x \rfloor$). It remains two kinds of $\handle_x$ to consider: \\

Subcase 1: Let $H_n(x)$ be a $\handle_x$ of $G$ of length $n \geq \Even(\frac{2}{x-\lfloor x \rfloor})$. Suppose that
$G'=G-\{v_1,\dots,v_{n-1}\}$ is $(a,b)$-choosable, and let $L$ be an $a$-list of $G$. Then there exists an  
$(L,b)$-choosability $c$ of $G'$. We set $N'(v_i)$ to be the set of neighbors of  $v_i$ in the induced subgraph 
$G-\{v_{i-1},v_{i+1}\}$. By hypothesis for all $i \in \{1,\dots,n-1\}$, the degree of $v_i$ satisfies $d(v_i) \leq \lfloor x \rfloor$ hence
$|N'(v_i)| \leq \lfloor x \rfloor -2$. Let $L'$ be the list of  $H_n(x)$ such that:
$$L'(i) = \left \{
\begin{array}{l}
c(v_i)  \ \  \ \ \ \  \ \ \ \ \ \ \  \ \ \  \ \ \ \  \  \ \ \  \ \ \  if \ i \in \{0,n\} \\
L(v_i) \setminus \bigcup_{w \in N'(v_i)} c(w) \ \ \  \  otherwise. 
\end{array}
\right.
$$ 
Then for all $i \in \{1,\dots,n-1\}$, we have:
$$|L'(i)| \geq |L(v_i)|- \sum_{w \in N'(v_i)}|c(w)| \geq a-(\lfloor x \rfloor -2)b=2b+e.$$
Suppose without loss of generality that for all  $i \in \{1,\dots,n-1\}$, we have: $|L'(i)|=2b+e$. 
Since $\frac{a}{b} \geq x$ we have $\frac{2}{x-\lfloor x \rfloor} \geq \frac{2b}{e}$ and thus
$$n \geq \Even(\frac{2}{x-\lfloor x \rfloor}) \geq \Even(\frac{2b}{e}).$$

Since $|L'(0)|=|L'(n)|=b$, then $L'$ is a list which satisfies the hypothesis of Corollary \ref{theorem48these}: we
obtain the existence of an $(L',b)$-choosability $c'$ of $P_{n+1}$, {\em i.e.} of $H_n(x)$. Finally, we construct an
$(L,b)$-choosability $c''$ of $G$ such that:

$$c''(v) = \left \{
\begin{array}{l}
c(v) \  \ \ \  if \ v \in G' \\
c'(v) \ \ \  \  otherwise. 
\end{array}
\right.
$$ 

Subcase 2: Let  $H_n(x)$ be a $1$-$\handle_x$ of $G$ of length $n = \Even(\frac{2}{x-\lfloor x \rfloor})-1$. If
$\Even(\frac{2}{x-\lfloor x \rfloor}) > \Even(\frac{2b}{e})$ then $n \geq \Even(\frac{2b}{e})$, it is a $\handle_x$ of
sufficiently big length, and hence we come back to Subcase 1.
Otherwise, 
 $\Even(\frac{2}{x-\lfloor x \rfloor}) = \Even(\frac{2b}{e})$ (because $\Even(\frac{2}{x-\lfloor x \rfloor}) \geq
\Even(\frac{2b}{e})$). By definition, there exists a vertex $v_{n+1} \in G - H_n (x)$ which is a neighbor of  $v_{n}$ such
that $d(v_{n+1}) \leq \lfloor x \rfloor$. Suppose that  $G'=G-\{v_1,\dots,v_{n-1}\}$ is $(a,b)$-choosable, and let $L $ be
an $a$-list of $G$ and $c$ be an $(L,b)$-choosability of $G'$. For any $i \in \{1,\dots,n\}$, $N'(v_i)$ is the set
of neighbors of $v_i$ in $G-\{v_{i-1},v_{i+1}\}$ and $N'(v_{n+1})$ is the set  of neighbors of $v_{n+1}$ in $G-\{v_{n}\}$. Let
$L'$ be the list of  $H_n(x)\cup \{v_{n+1}\}$ such that:
$$L'(i) = \left \{
\begin{array}{l}
c(v_0)  \ \  \ \ \ \  \ \ \ \ \ \ \  \ \ \  \ \ \ \  \  \ \ \  \ \ \  if \ i =0 \\
L(v_i) \setminus \bigcup_{w \in N'(v_i)} c(w) \ \ \  \  otherwise. 
\end{array}
\right.
$$ 
For  any $i \in \{1,\dots,n-1\}$, we have:
$$|L'(i)| \geq |L(v_i)|- \sum_{w \in N'(v_i)}|c(w)| \geq a-(\lfloor x \rfloor -2)b=2b+e.$$
Furthermore, for $i \in \{n,n+1\}:$
$$|L'(i)| \geq |L(v_{i})|- \sum_{w \in N'(v_{i})}|c(w)| \geq a-(\lfloor x \rfloor -1)b=b+e.$$
We can suppose without loss of generality that for all $i \in \{1,\dots,n-1\}$, we have $|L'(i)|=2b+e$, and for $i \in \{n,n+1\}$, 
we have $|L'(i)|=b+e$. Since $c$ is an $(L,b)$-choosability of  $G'$ then $c(v_{n}) \cap c(v_{n+1})=\emptyset$ and  
since for $i \in \{n,n+1\}$, we have $c(v_i) \subset L'(i) $ then we obtain:
$$ |A(n,n+1)(L')|=|L'(n) \cup L'(n+1)| \geq |c(v_{n})| + |c(v_{n+1})|=2b.$$
Hence $L'$ is a $1$-reduced list of $H_n(x)\cup \{v_{n+1}\}$;  but since $n+1=\Even(\frac{2b}{e})$, Theorem \ref{theorem49these}
constructs an $(L',b)$-choosability $c'$ of $H_n(x)\cup \{v_{n+1}\}$. Finally, we construct an $(L,b)$-choosability $c''$ of $G$ such
that:

$$c''(w) = \left \{
\begin{array}{l}
c(w) \  \ \ \  if \ w \in G - (H_n(x)\cup \{v_{n+1}\}) \\
c'(w) \ \ \  \  otherwise. 
\end{array}
\right.
$$

In conclusion, if $\Core_{ch}(x,1)(G)$ is $(a,b)$-choosable, we add successively the subgraphs of $G$ that we removed (to
obtain the $\Core_{ch}(x,1)$), in the opposite order. Thanks to the cases studied in every step, the graph remains
$(a,b)$-choosable and thus $G$ is $(a,b)$-choosable.
\end{proof}

Let us define the set
 $\mathcal{C}_h(x)$ to be the set of graphs $G$ which are $(a,b)$-choosable for all $a,b$ such that $\frac{a}{b} \geq x$, {\em i.e.}:

$$\mathcal{C}_h(x)=\{ G , \text{such that for all} \ \frac{a}{b} \geq x, \ G \ \text{is} \ (a,b)\text{-choosable}\}.$$

Hence, we deduce the following corollary:

\begin{corollary}
\label{corollaire6these}
Let $G$ be a graph and $x$ be a rational number. Then:
$$G \in \mathcal{C}_h(x) \Leftrightarrow \Core_{ch}(x,1)(G) \in \mathcal{C}_h(x).$$
\end{corollary}





\section{Application to the triangular lattice}

Let $\mathcal{R}$ be a triangle-free induced subgraph of the triangular lattice.
Recall that the triangular lattice is embedded in an Euclidian space and that any vertex $(x,y)$ of $\mathcal{R}$ has at most six neighbors: 
{its neighbor on the left},  $(x-1,y)$, {its neighbor on the right} $(x+1,y)$, {its neighbor on the top left}
$(x-1,y+1)$, {its neighbor on the top right} $(x,y+1)$, {its neighbor on the bottom left}
$(x,y-1)$ and {its neighbor on the bottom right} $(x+1,y-1)$.\\

\begin{definition}
The nodes of $\mathcal{R}$ are the vertices of degree  $3$. There are two kinds of nodes: the left nodes whose neighbors
are the neighbors on the left, on the top right, and on the bottom right; and the right nodes whose
neighbors are the neighbors on the right, on the top left and on the bottom left.
\end{definition}

\unitlength=0.6cm
\begin{picture}(20,4)(0,0)
\put(6.5,2.5){\circle*{0.3}}
\put(7.5,4){\circle*{0.3}}
\put(7.5,1){\circle*{0.3}}
\put(5,2.5){\circle*{0.3}}
\put(13.5,2.5){\circle*{0.3}}
\put(15,2.5){\circle*{0.3}}
\put(12.5,1){\circle*{0.3}}
\put(12.5,4){\circle*{0.3}}
\put(5,2.5){\line(1,0){1.5}}
\put(6.5,2.5){\line(2,3){1.1}}
\put(6.5,2.5){\line(2,-3){1.1}}
\put(13.5,2.5){\line(1,0){1.5}}
\put(13.5,2.5){\line(-2,3){1.1}}
\put(13.5,2.5){\line(-2,-3){1.1}}
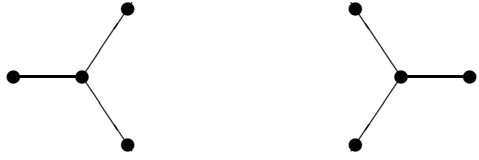
\put(4,.1){Fig. 4. Left node and right node. }
\end{picture}

\begin{definition}
A cutting node of  $\mathcal{R}$ is a left node   $(x,y)$ such that for any node   $(x',y')$, we have $y \geq y'$, and
for any left node  $(x',y)$, we have $x' \leq x$. A handle of   $\mathcal{R}$ is a path  $P$ such that its extremal
vertices  are nodes and its internal vertices have degree 2.

A cutting handle of $\mathcal{R}$  is a handle such that one of its extremal vertices is the cutting node $(x,y)$ and
one of its internal vertices is $(x,y+1)$. 
\end{definition}

One can see the cutting node as the left node the most on the top on the right.

We have the trivial following lemma:

\begin{lemma}
\label{lemme22these}
Let $P_{n+1}$ be a cutting handle of  $\mathcal{R}$ such that $V(P_{n+1})=\{v_0,\dots,v_n\}$. If the length   $n$ of
$P_{n+1}$ is less or equal to  $3$, then $n=3$ and $v_{3}$ has got a neighbor  $v_4 \not= v_{2} $ of degree less or
equal to  $2$.
\end{lemma}

\unitlength=0.7cm
\begin{picture}(20,5)(0,0)
\put(6.5,2.5){\circle*{0.3}}
\put(7.5,4){\circle*{0.3}}
\put(9,4){\circle*{0.3}}
\put(10,2.5){\circle*{0.3}}
\put(11.5,2.5){\circle*{0.3}}
\put(9,1){\circle*{0.3}}
\put(7.5,1){\circle*{0.3}}
\put(5,2.5){\circle*{0.3}}
\put(6.5,2.5){\line(2,3){1}}
\put(6.5,2.5){\line(2,-3){1}}
\put(7.5,4){\line(1,0){1.5}}
\put(7.5,1){\line(1,0){1.5}}
\put(10,2.5){\line(1,0){1.5}}
\put(9,1){\line(2,3){1}}
\put(10,2.5){\line(-2,3){1}}
\put(5,2.5){\line(1,0){1.5}}
\put(6.2,2.9){$v_{0}$}
\put(7.2,4.4){$v_{1}$}
\put(8.8,4.4){$v_{2}$}
\put(9.9,2.9){$v_{3}$}
\put(11.4,2.9){$v_{4}$}
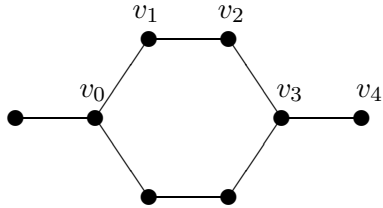
\put(3,.1){Fig. 5. Cutting handle of length $3$.}
\end{picture}

We have:

\begin{theorem}
\label{theoreme55these}If $\mathcal{R}$ is a triangle-free induced subgraph of the triangular lattice, then:
$$\mathcal{R} \in \mathcal{C}_h(\frac{5}{2}) \ .$$
\end{theorem}

\begin{proof}
We set $x=\frac{5}{2}$, thus $\lfloor x \rfloor = 2$, hence a handle of  $\mathcal{R}$ is a  $\handle_x$ of
$\mathcal{R}$. Let $G=\Core_{ch}(x,1)(\mathcal{R})$. If $G = \emptyset$ then Corollary   \ref{corollaire6these} gives the
result.  Otherwise $G \not= \emptyset$, and since the girth of this graph is at least   $6$ (and since we have
removed all the   $\handle_x$ of length $n \geq \Even(\frac{2}{x-\lfloor x \rfloor })=4$), $G$ can't be a forest of
cycles and thus $G$ has got at least two nodes (since we have removed the vertices of degree 0 and 1, the number of
nodes is necessarily even). By symmetry, one can suppose that $G$ has a cutting handle (otherwise we can consider its
mirror graph). Since $x=\frac{5}{2}$ and since we have removed all the $\handle_x$ of length $n \geq 4$, then Lemma
\ref{lemme22these} shows that this cutting handle $P_{n+1}$ has length $n=3$, $V(P_3)=\{v_0,\dots,v_3\}$ and $v_3$ has a
neighbor $v_4 \not=v_2$ of degree less or equal to $2$. Thus $P_3$ is a $1$-$\handle_x$, but this is absurd since
$G=\Core_{ch}(x,1)(\mathcal{R})$. 
\end{proof}





\section{The second extended core}

We can define a second extended core by removing more vertices.

\begin{definition}
A $2$-$\handle_x$ of length $n$ in a graph $G$ is a $1$-$\handle_x$ of length $n$ such that $v_0$ has degree less or equal to   $\lfloor x +1 \rfloor$
and has a neighbor  $v_{-1}$ of degree less or equal to  $\lfloor x \rfloor$ and for all $i,j \in \{-1,\dots,n+1\}$ we
have: if $|i-j| \geq 2$ then $v_i v_j \notin E$. 
\end{definition}

\unitlength=0.8cm
\begin{picture}(20,4.5)
\put(3,2){\line(1,0){10}}
\put(3,2){\circle*{0.2}}
\put(4,2){\circle*{0.2}}
\put(5,2){\circle*{0.2}}
\put(11,2){\circle*{0.2}}
\put(12,2){\circle*{0.2}}
\put(13,2){\circle*{0.2}}
\put(14,3){\circle*{0.2}}
\put(2,3){\circle*{0.2}}
\put(2.8,1.2){$v_0$}
\put(1.4,2.2){$v_{-1}$}
\put(3.8,1.2){$v_1$}
\put(4.8,1.2){$v_2$}
\put(10.8,1.2){$v_{n-2}$}
\put(11.8,1.2){$v_{n-1}$}
\put(12.8,1.2){$v_n$}
\put(13.8,2.2){$v_{n+1}$}
\put(3.6,2.8){$\lfloor x \rfloor$}
\put(4.6,2.8){$\lfloor x \rfloor$}
\put(10.6,2.8){$\lfloor x \rfloor$}
\put(11.6,2.8){$\lfloor x \rfloor$}
\put(13.6,.8){$\lfloor x+1 \rfloor $}
\put(13.6,3.8){$\lfloor x \rfloor$}
\put(1.6,3.8){$\lfloor x \rfloor$}
\put(.8,.8){$\lfloor x+1 \rfloor $}
\put(2.7,1.7){\line(1,1){.6}}
\put(3.7,1.7){\line(1,1){.6}}
\put(4.7,1.7){\line(1,1){.6}}
\put(10.7,1.7){\line(1,1){.6}}
\put(11.7,1.7){\line(1,1){.6}}
\put(12.7,1.7){\line(1,1){.6}}
\put(3.3,1.7){\line(-1,1){.6}}
\put(4.3,1.7){\line(-1,1){.6}}
\put(5.3,1.7){\line(-1,1){.6}}
\put(11.3,1.7){\line(-1,1){.6}}
\put(12.3,1.7){\line(-1,1){.6}}
\put(13.3,1.7){\line(-1,1){.6}}
\multiput(6,1.4)(.3,0){15}{\line(1,0){0.05}}
\put(13,2){\line(1,1){1.3}}
\put(14.3,2.7){\line(-1,1){.6}}
\put(3,2){\line(-1,1){1.3}}
\put(1.7,2.7){\line(1,1){.6}}
\put(4,.1){Fig. 6. Example of a $2$-$\handle_x$.}
\end{picture}

\bigskip

Then,  we define the second extended core, denoted  $\Core_{ch}(x,2)(G)$, of a graph $G$ as the induced subgraph of $G$ obtained inductively when we remove:
\begin{itemize}
\item its vertices of degree  $0,1,\dots,\lfloor x-1 \rfloor$, 
\item the interior of its  $\handle_x$ of length $n \geq \Even(\frac{2}{x-\lfloor x \rfloor})$,
\item  the interior of its $1$-$\handle_x$ of length $n = \Even(\frac{2}{x-\lfloor x \rfloor}) -1$,
\item  the interior of its $2$-$\handle_x$ of length $n = \Even(\frac{2}{x-\lfloor x \rfloor}) -2$.
\end{itemize}

We can prove (see \cite{god09} for the details) in the same way as for the first extended core that the choosability of a graph 
reduces to the choosability of its second extended core:

\begin{theorem}
\label{theoremcorech2}
Let $a,b$ be two integers and $x$ be a rational number such that   $\frac{a}{b} \geq x$. For any graph $G$, we have the following equivalence:

$$G \ is \ (a,b)-choosable \Leftrightarrow \Core_{ch}(x,2)(G) \ is \ (a,b)-choosable.$$
\end{theorem}

\begin{definition}
Let $x \in [2,3[$ be a rational number. A $\handle_x$ of parity of a graph  $G$ is a  $\handle_x$ of length  $n$, 
such that there exists another path  in $G$ of length $m$, from $v_0$ to $v_n$, with $n \geq m$ and $m \equiv n \pmod{2}$.
\end{definition}

\begin{definition}
The  $\Core_{co}(x,2)(G)$ is defined as the  $\Core_{ch}(x,2)(G)$, but in addition, we remove also, for $x \in [2,3[$, the
interior of its  $\handle_x$ of parity.
\end{definition}

Let us define the set
 $\mathcal{C}_o(x)$ to be the set of graphs $G$ which are $(a,b)$-colorable for all $a,b$ such that $\frac{a}{b} \geq x$, {\em i.e.}:

$$\mathcal{C}_o(x)=\{ G , \text{such that for all} \ \frac{a}{b} \geq x, \ G \ \text{is} \ (a,b)\text{-colorable}\}.$$

The McDiarmid and Reed conjecture  (see \cite{mcd00}) asserts that:
$$\mathcal{R} \in \mathcal{C}_o(\frac{9}{4}) \ .$$

One can prove in the same way as for the  $\Core_{ch}(x,2)(G)$ (see   \cite{god09} for the details) the following theorem:

\begin{theorem} 
\label{corollaire10these}
Let $x \in [2,3[$. For any graph $G$, we have:
$$ G \in \mathcal{C}_o(x) \Leftrightarrow \Core_{co}(x,2)(G) \in \mathcal{C}_o(x).$$
\end{theorem}

Then, Theorem \ref{corollaire10these} enables us  to find again a result of Havet on the coloration of the graph
$\mathcal{R}$:

\begin{theorem}[Havet, \cite{hav01}]
\label{havet73}
$$\mathcal{R} \in \mathcal{C}_o(\frac{7}{3}) \ .$$
\end{theorem}

\begin{proof}
We set $x=\frac{7}{3}$, and $G=\Core_{co}(x,2)(\mathcal{R})$. If $G=\emptyset$ then Theorem \ref{corollaire10these} gives
the proof. Otherwise, $G \not= \emptyset$ and there exists a cutting handle $P_{n+1}$ of $G$. Since
$\Even(\frac{2}{x-\lfloor x \rfloor})=6$, we have removed all the $\handle_x$ of length $n \geq 6$. For $n\leq 5$, a
careful study (see \cite{god09} where some technical lemmas are used) shows that we obtain a contradiction, which
concludes the proof. 
\end{proof}

\begin{remark}
Using the $\Core_{ch}(7/3,2)$, we can almost prove (some few cases still resist) that $\mathcal{R} \in \mathcal{C}_h(\frac{7}{3})$.
\end{remark}

\begin{remark}
Recall that a graph $G$ is said to be $(a,b)$-free-choosable in a vertex $v_0$ if for any  list $L$ of $G$ such that for any 
$v \in V(G) \setminus \{v_0\}:|L(v)|=a$ and $|L(v_0)|=b$, there exists an  $(L,b)$-choosability $c$ of $G$.
If $G$ is a graph such that $\Core_{ch}(x,2)(G)=\emptyset$, then we have shown in particular that if $a,b$ are such that 
$\frac ab \geq x$ then $G$ is $(a,b)$-free-choosable in the last vertex removed by the algorithm used to get the core. 
In particular, Theorem \ref{theoreme55these} implies that there exists a vertex $v_0$ for which the graph $\mathcal{R}$ 
is $(5,2)$-free-choosable in $v_0$.
\end{remark}






\begin{thebibliography}{3}

\bibitem{nous} Y. Aubry, J.-C. Godin, O. Togni,
\newblock {\sl Choosability of a weighted path and free-choosability of a cycle},
\newblock arXiv:1005.5602v1 [math.CO] (2010).

\bibitem{borodinkostochkawoodall97} O.V. Borodin, A.V. Kostochka, D.R. Woodall,
\newblock {\sl List edge and list colourings of multigraph},
\newblock J. Combin. Theory Series B, 71 : 184-204, (1997).


\bibitem{erdo79} P. Erd\H{o}s, A.L Rubin and H. Taylor,
\newblock {\sl Choosability in graphs},
\newblock Proc. West-Coast Conf. on Combinatorics, Graph Theory and Computing, Congressus Numerantium XXVI, (1979), 125-157.


\bibitem{god09} J.-C. Godin,
\newblock {\sl Coloration et choisissabilit\'e des graphes et applications},
\newblock PhD thesis (in french), Universit\'e du Sud Toulon-Var, France (2009).

\bibitem{gravier96} S. Gravier,
\newblock {\sl A Haj\'os-like theorem for list coloring},
\newblock Discrete Math. 152, (1996), 299-302.


\bibitem{GutnerTarsi} S. Gutner and M. Tarsi,
\newblock {\sl Some results on (a:b)-choosability},
\newblock Discrete Math. 309, (2009), 2260-2270.

\bibitem{hav01} F. Havet,
\newblock {\sl Channel assignement and multicolouring of the induced subgraphs of the triangular lattice}.
\newblock Discrete Math. 233, (2001), 219-233.

\bibitem{hav09} F. Havet,
\newblock {\sl Choosability of the square of planar subcubic graphs with large girth},
\newblock Discrete Math. 309, (2009), 3553-3563.

\bibitem{mcd00} C. McDiarmid and B. Reed,
\newblock {\sl Channel assignement and weighted coloring}.
\newblock Networks, 36, (2000), 114-117.

\bibitem{tom94} C. Thomassen
\newblock {\sl Every planar graph is $5$-choosable}.
\newblock J. Combin. Theory Ser. B 62 (1994), no. 1, 180--181. 

\bibitem{tuz96} Zs. Tuza and M. Voigt,
\newblock {\sl Every 2-choosable graph is (2m,m)-choosable},
\newblock J.  Graph Theory 22, (1996), 245-252.

\bibitem{viz76} V. G Vizing,
\newblock {\sl Coloring the vertices of a graph in prescribed colors (in Russian)},
\newblock Diskret. Analiz. No. 29, Metody Diskret. Anal. v Teorii Kodov i Shem 101 (1976), 3-10.

\bibitem{voi96} M. Voigt,
\newblock {\sl Choosability of planar graphs},
\newblock Discrete Math., 150, (1996), 457-460.  

\bibitem{voi98} M. Voigt,
\newblock {\sl On list Colourings and Choosability of Graphs},
\newblock Abilitationsschrift, TU Ilmenau (1998). 

\end{thebibliography}
\end{document}